\documentclass[10pt, a4paper]{article}
\usepackage{graphicx}
\usepackage[table,xcdraw]{xcolor}
\usepackage{listings}
\usepackage{amssymb,amsmath,amsthm}
\usepackage{tabularx}
\usepackage{makecell}
\usepackage{float}
\usepackage{ragged2e}
\usepackage{booktabs}
\usepackage{array}
\usepackage{xurl}
\usepackage{float}
\usepackage{placeins}
\usepackage[utf8]{inputenc}
\usepackage{geometry}
\usepackage{titling}
\usepackage[hidelinks]{hyperref}
\usepackage{tikz}
\usetikzlibrary{positioning, shapes.geometric, arrows}
\usepackage{pgfplots}
\pgfplotsset{compat=newest}

\pretitle{
  \begin{center}
  \large
  \vskip 0.4em
  \hrule height 3pt
  \vspace{0.1cm}
  \LARGE\bfseries
}
\posttitle{
  \vskip 0.3cm
  \hrule height 1pt
  \vskip 0.5em
  \end{center}
}
\preauthor{\large\begin{center}}
\postauthor{\end{center}}
\predate{\large\begin{center}}
\postdate{\end{center}}

\begin{document}

\title{BasedAI: A decentralized P2P network for Zero Knowledge Large Language Models (ZK-LLMs)}
\preauthor{\begin{center}\small Version 1.0 \\[1em]}
\author{\textbf{Sean Wellington}}
\postauthor{\\[1em] Based Labs\end{center}}

\date{\footnotesize February 29th, 2024}

\maketitle
\begingroup
\renewcommand{\thefootnote}{}
\footnote{\url{nnlib2ju6byq6hj4jgmcamcfulyaji5zeptspskcbn6zw4j6nsbm4dad.onion/wp.pdf} | \url{https://getbased.ai}}
\addtocounter{footnote}{-1}
\endgroup
\begin{center}
\section*{Abstract}
\end{center}

BasedAI is a distributed network of machines which introduces decentralized infrastructure capable of integrating Fully Homomorphic Encryption (FHE) with any large language model (LLM) connected to its network. The proposed framework embeds a default mechanism, called “Cerberus Squeezing”, into the mining process which enables the transformation of a standard LLMs into encrypted zero-knowledge LLMs, or “ZK-LLMs”, leveraging insights from generative adversarial networks for data privacy (Goodfellow, I. J. et al. 2014\cite{goodfellow2014}). This novel quantization mechanism empowers BasedAI miners to process and respond to prompts derived from User interaction with LLMs without the need for decrypting either the queries or their corresponding responses. The introduction of Cerberus Squeezing significantly improves performance degradation caused by quantized functions in current FHE-compliant computing environments by proactively optimizing calls between users, miners, and validators.

While this paper primarily examines the application of BasedAI within the realm of LLMs, it is important to note that the underlying architecture of BasedAI is inherently versatile, with potential for expansion into other domains. The core contribution of this work lies in addressing the challenge of maintaining privacy, while efficiently executing complex computations, achieved through BasedAI’s peer-to-peer network structure.

\newpage
\tableofcontents

\section{Introduction}

The widespread deployment of large language models (LLMs) in critical domains has underscored a pressing need for frameworks that ensure data privacy without compromising computational performance. Fully Homomorphic Encryption (FHE) appears as a solution to this need, providing the capability to perform calculations on encrypted data. However, the computational burden posed by FHE, when combined with the resource-intensive nature of LLMs, poses a significant challenge to preserving both privacy and service quality in distributed AI systems.

In response to this challenge, BasedAI proposes a decentralized approach for seamlessly integrating FHE with LLMs to maintain data confidentiality without significant performance trade-offs (Gentry, C. 2009 \cite{gentry2009}). This paper outlines the BasedAI architecture and introduces the novel Cerberus Squeezing technique aimed at enhancing the efficiency of encrypted computations. As we examine the technical components and potential applications of BasedAI further, we aim to demonstrate that it is possible to reconcile the seemingly conflicting demands of data security and processing power within a decentralized, privacy-preserving computational model.

\section{BasedAI: Decentralized Network Architecture}

\subsection{Embedding Privacy at the Core}
The advent of privacy-centric cryptocurrencies like Zcash and Monero marked a significant advancement in digital privacy, integrating advanced cryptographic techniques such as zk-SNARKs and ring signatures to secure transactions while ensuring user anonymity (Bitansky, N. 2013 \cite{bitansky2013}). Their emergence not only underscored a growing demand for enhanced privacy but also fundamentally altered the landscape of financial transactions in the digital domain, demonstrating the critical need for and effectiveness of privacy in the digital age.

Mirroring the ethos of these projects, BasedAI embeds privacy at the core of its distributed network infrastructure with Zero-Knowledge Large Language Models (ZK-LLMs). The protocol employs innovative techniques to ensure user data remains private throughout the network's operations, effectively handling sensitive information without sacrificing computational efficacy. By doing so, BasedAI presents a far more data-protected option for both network infrastructure and end-users.
\subsection{The BasedAI Platform}

BasedAI's platform operates on a decentralized network primarily composed of entities known as "Brains." These Brains act as distributed containers for specific computational tasks, mainly for running modified large language models (LLMs). Each Brain can select the LLM it wants its associated miners and validators to operate. BasedAI has the capability to transform any LLM operating on a Brain into a Zero-Knowledge Large Language Model (ZK-LLM). It does this by pairing Fully Homomorphic Encryption (FHE) with BasedAI’s quantization process (Cerberus Squeezing) ensuring data remains encrypted during both processing and delivery.

Owning a Brain can be compared to owning a cloud services license in BasedAI; owners do not have to use their Brains but are incentivized to do so through \$BASED tokens, which are distributed to Brain owners, giving relatively more \$BASED to high-performing Brains. These tokens are used by Brain owners to encourage the delegation of computational resources by miners and validators, who in turn receive rewards for their contributions.

The BasedAI network supports a capped quantity of Brains, along with a finite number of Validators and Miners, creating a competitive ecosystem. All participants are motivated through a rewards system where compensation is aligned with their level of contribution and the efficiency they bring to the network. This system, underpinned by \$BASED token rewards, propels all parties involved to establish and uphold high-performance models.

This platform was developed by Based Labs, in collaboration with the founding team of Pepecoin. Pepecoin (\url{https://etherscan.io/token/0xA9E8aCf069C58aEc8825542845Fd754e41a9489A}) plays a dual role within the BasedAI network, functioning both as currency for acquiring Brains and as the governing medium for network-wide decisions.

\subsection{Participant Roles in the BasedAI Ecosystem}

\subsubsection{Brain Owners}

\begin{figure}[h]
  \centering
  \includegraphics[width=0.1\linewidth]{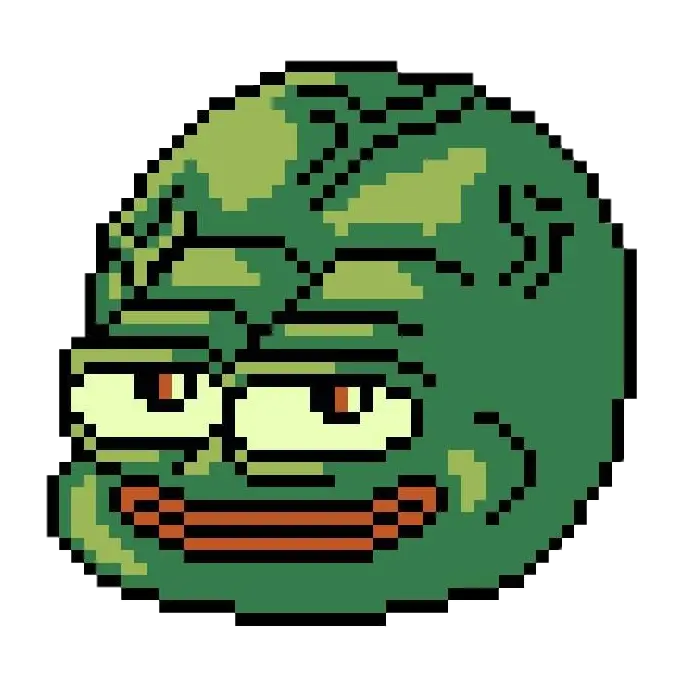}
  \caption{A “Brain” from Based Labs}
  \label{fig:brains}
\end{figure}

Brain Owners are incentivized with \$BASED to manage their Brains properly by selecting worthwhile ZK-LLMs and incentive models that will attract computational power from both miners and validators. Additionally, the more \$BASED staked to a Brain and the more efficient Miners and Validators are at running on their Brain, the more \$BASED rewards they all earn. To become a Brain Owner, you must acquire one via Pepecoin (more in Section 2.4.1) or the open market.

\subsubsection{Miners}

Miners are incentivized to pick a Brain that is high performing but does not already have a high number of miners, to run the model required by the given Brain, and to report truthfully. GPUs are preferable for these tasks as the computational resource requirements for running LLMs are higher. To become a miner:

\begin{enumerate}
    \item Pick a Brain
    \item Register your wallet with the Brain
    \item Choose the option to dedicate GPU resources
    \item Choose how much of your device's storage to allocate
    \item Optional: Stake your \$BASED to the Brain
    \item Start Mining
\end{enumerate}

\subsubsection{Validators}

Validators are incentivized to pick a Brain that is high-performing yet does not already have a large number of validators, to ensure that miners are correctly running the model required by their selected Brain, and to report truthfully. CPU machines are preferable as validators, as the computational resources required are lower to perform this task. To become a validator:

\begin{enumerate}
    \item Pick a Brain
    \item Register your wallet with the Brain
    \item Choose the option to dedicate CPU resources
    \item Choose how much of your device's storage to allocate
    \item Optional: Stake your \$BASED to the Brain
    \item Start Validating
\end{enumerate}

\subsubsection{Brain Permanent Memory}

To become a validator or a miner you must run a permanent memory operation to have that Brain “memorize” your validator or miner address. The memorize function may have a fee depending on the Brain owner’s preferences.

\subsection{Brain Dynamics and Operations}
\subsubsection{Acquiring Brains}
Brains can be acquired using Pepecoin in two different methods; burning and staking. Both methods result in the issuance of an ERC-721 token, which represents ownership of the Brain. The total issued supply is capped at 1024 Brains. A simple BasedAI portal will be available for users to execute their preferred method and to link their ETH wallet to a new Based wallet:

\paragraph{Burn Method:}
\begin{itemize}
    \item Requirement: User burns 1,000 Pepecoin
    \begin{itemize}
      \item This number increases by 200 Pepecoin per Brain issued via the Burn method
    \end{itemize}
    \item The Brain ERC-721 is issued immediately upon burn
    \item Brain is transferrable
\end{itemize}

Note: If all Brains are acquired through the Burn method, 107,563,530 Pepecoin will be permanently burned.

\begin{figure}[H]
  \centering
  \includegraphics[width=.75\linewidth]{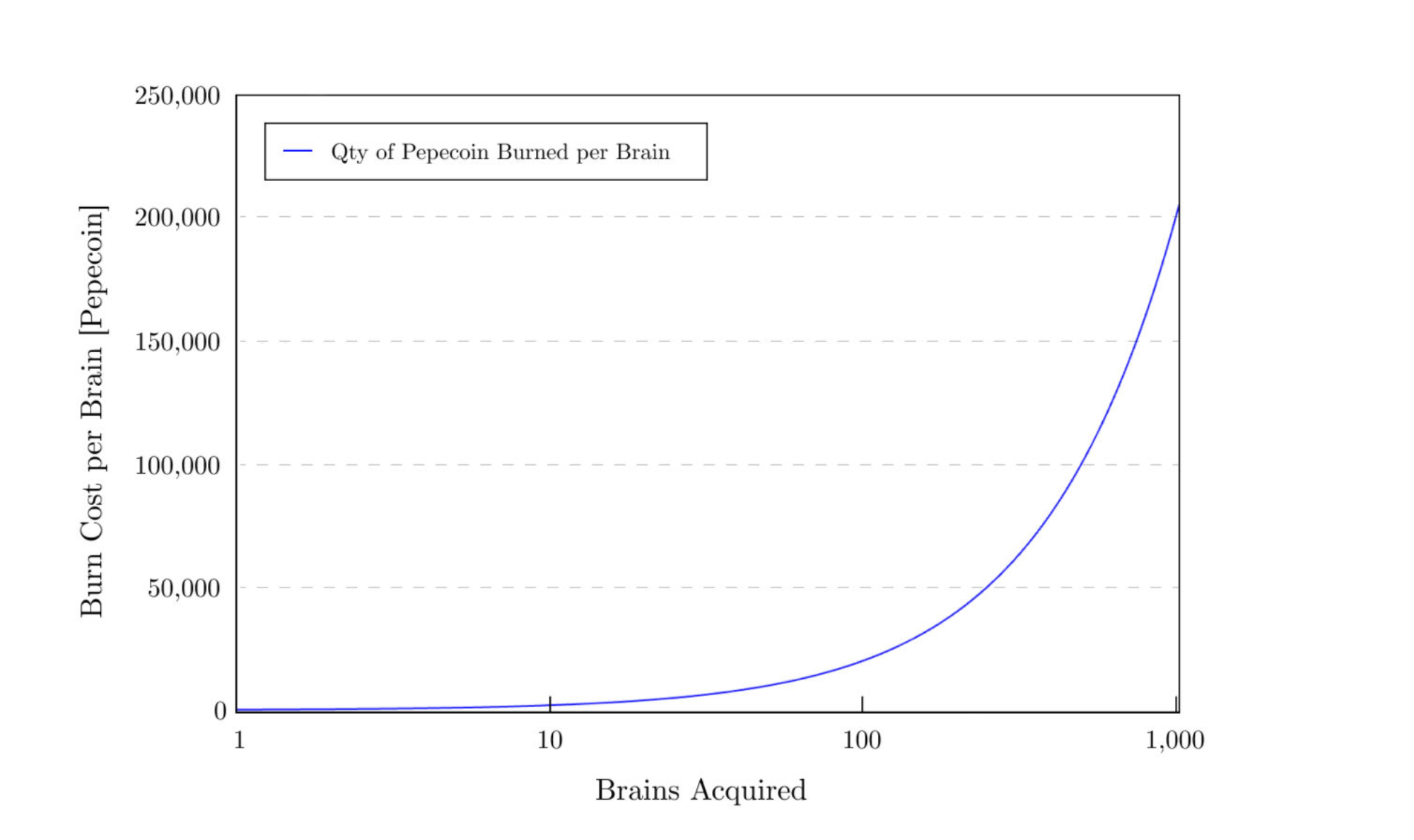}
  \caption{Pepecoin cost per Brain increases with each burn}
  \label{fig:burnPepecoin}
\end{figure}

\paragraph{Stake Method:}
\begin{itemize}
    \item Requirement: User stakes 100,000 Pepecoin for a period of 90 days
    \item The Brain ERC-721 is issued immediately upon deposit
    \item Brain is non-transferrable
    \item After the 90 days:
    \begin{itemize}
      \item The Based wallet associated to the Brain begins accumulating \$BASED rewards
      \item Pepecoin can be unstaked by disabling the Brain (discussed in 2.4.2)
    \end{itemize}
\end{itemize}

Note: This is not the same process as staking \$BASED to a Brain, which earns users \$BASED rewards. Staking Pepecoin via the BasedAI Portal is a one time process used to acquire a Brain NFT.

As more Brains are created, a commensurate amount of Pepecoin is either burned or locked up, depending on the ratio of participation in the two methods.

\begin{figure}[H]
  \centering
  \includegraphics[width=.75\linewidth]{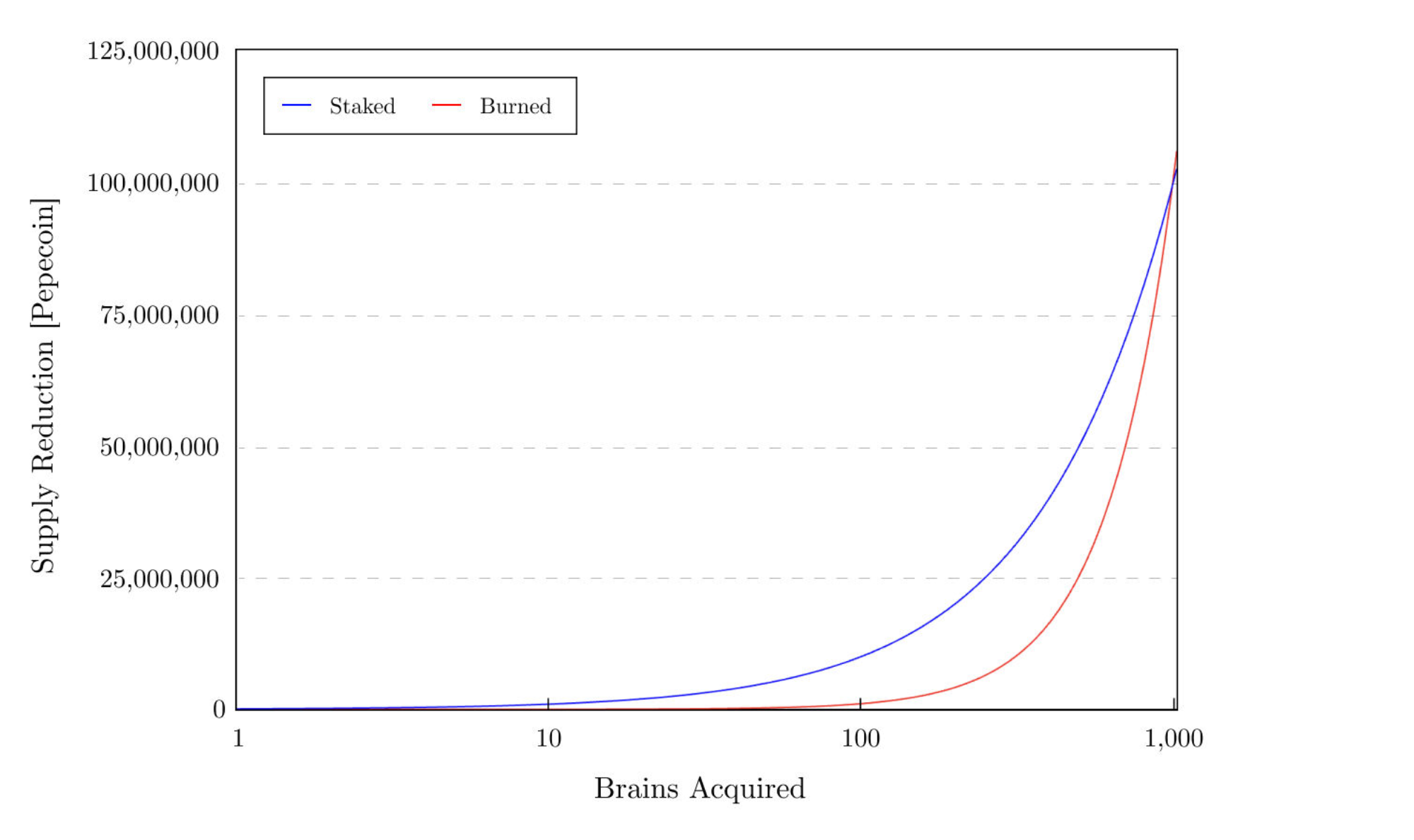}
  \caption{Pepecoin supply potentially burned or staked in the Brain creation process}
  \label{fig:stakePepecoin}
\end{figure}

As long as less than 1024 Brains are issued and active in the ERC-721 contract, the BasedAI Portal will continue to issue Brains. If all 1024 Brains are issued, the BasedAI Portal will not permit the creation of new Brains.
Multiple Brain NFTs can be held at one Ethereum address. The BasedAI Portal will allow users to manage their rewards earned from all owned Brains associated with a connected ETH wallet. Active Brain owners are estimated to earn between 30,000 and 80,000 \$BASED annually per Brain.

\subsubsection{Brain Deactivation Procedure}

Only Brains acquired via Staking can be deactivated. The deactivation process requires the return of the ERC-721 token to the Brain contract, this will unlock the owner’s staked Pepecoin. Once a Brain is deactivated, the available supply of Brains that can be acquired via the BasedAI Portal goes up by one.

\subsubsection{Operating and Managing a Brain}
The administration of Brains under the BasedAI network is facilitated through an interface accessible at \url{https://www.getbased.ai}. Brain owners are provided with several management features to tailor the operational aspects of their Brains effectively. Available options include:

\begin{itemize}
    \item Selection of the computational task (LLM) to be executed by miners and validators within the network.
    \begin{itemize}
      \item Note: Changing the computational task at a Brain is a trivial process. Miners and validators will be given notice and will simply restart with the code provided.
    \end{itemize}
    \item Allocation ratio of \$BASED token emissions for both miners and validators.
    \item Configuration of the Brain's identity parameters, such as its name and its ENS/BNS designation.
    \item Specification of the ss58 address for receiving reward distributions.
\end{itemize}

\subsubsection{Build Your Own Brain Templates}

Templates from Based Labs will be provided to Brain owners to assist with the launch process. These templates issue ready-to-run code for validators and miners who wish to begin working on that Brain. Although it is possible to run various tasks, BasedAI is specifically optimized to run ZK-LLMs.

The first Brain template made available on BasedAI mainnet is the Hugging Face ZK-LLM Converter. This template allows Brain owners to specify a Hugging Face URL for a given model and automatically import that model to their Brain as the computational task for the miners and validators, all with ZK-LLM functionality. As mentioned earlier in this paper, ZK-LLM functionality is highly experimental, but to drive innovation and education around the importance of private anonymous artificial intelligence, it is enabled by default but can be disabled.

\subsubsection{Administrative Brains Operated by Based Labs}
All Brains issued will be numbered. There will be eight Brains in BasedAI (\#0 - \#6 and \#47) that have specific roles and will be maintained by Based Labs.  Brains \#0 through \#2 are administrative, with Brains \#3 and \#4 are for demonstrating high performance industry applications in a ZK environment and Brains \#5, \#6, and \#47 are reserved for upcoming BasedAI capabilities.
\renewcommand{\arraystretch}{1.5}

\begin{table}[ht]
\centering
\small
\begin{tabularx}{\textwidth}{|X|X|X|}
\hline
\multicolumn{1}{|c|}{Brain Number} & \multicolumn{1}{c|}{Function} & \multicolumn{1}{c|}{Additional Info} \\
\hline
Brain \#0 & Distributes \$BASED rewards & 10 \$BASED per block; Blocks issued every 10s. \\
Brain \#1 & Operates BasedEVM & High-performance smart contract layer on Rust. \\
Brain \#2 & Houses the TFT Enforcer Model. & Will become Brain \#0 (Section 3.4.2) \\
Brain \#3 & Dedicated to Native ZK-LLM. & Approximate size of \~15GB. \\
Brain \#4 & Dedicated to ZK-Video. & Private encrypted Video generation. \\
Brain \#5 & Reserved for Based Labs & Allocated for future use. \\
Brain \#6 & Reserved for Based Labs & Allocated for future use. \\
Brain \#47 & Reserved for Based Labs & Allocated for future use. \\
\hline
\end{tabularx}
\caption{Roles of the eight Brains operated by Based Labs}
\label{tab:brainRoles}
\end{table}

\renewcommand{\arraystretch}{1}

\section{Tokenomics, Governance, and Utility in BasedAI}

\subsection{Rewarding Participation: Incentives and Tokens}
\subsubsection{Reward Distribution}

BasedAI issues 10 \$BASED tokens every 10 seconds as an incentive to Brains within the network. This distribution is made proportionally, based on two criteria: the stake volume of \$BASED tokens invested in each Brain and the consistent operational performance demonstrated by the miners and validators associated with those Brains. By meticulously balancing these factors, the protocol ensures that rewards correlate directly with active, stable, and substantial participation in the network's ecosystem. The most proficient Brains—those in the highest-performing 30\%—are eligible for additional \$BASED bonuses.

Any \$BASED tokens emitted by the network are two-way bridgeable with the wrapped BasedAI ERC-20 token (\url{https://etherscan.io/token/0x44971abf0251958492fee97da3e5c5ada88b9185}).

\subsubsection{Halving Schedule}

To manage inflationary risks, \$BASED's emission schedule undergoes a halving event annually. The halving also serves to sustain mining incentives over time, encouraging long-term commitment and investment from network participants through this deflationary mechanism.

Below is the detailed schedule overviewing how the block reward and subsequent token emission rates are set to decrease beginning April 2024.

\FloatBarrier
\begin{table}[H]
\small
\centering
\begin{tabularx}{\textwidth}{
    |>{\hsize=0.5\hsize\centering\arraybackslash}X|
    >{\hsize=0.5\hsize\raggedleft\arraybackslash}X|
    >{\hsize=1\hsize\raggedleft\arraybackslash}X|
    >{\hsize=1\hsize\raggedleft\arraybackslash}X|
}
\hline
Year & Block Reward & \$BASED Emission & TFT Enforcer Weight \\
\hline
1 & 10 & 31,536,000.00 & 0 \\
2 & 5 & 15,768,000.00 & 0 \\
3 & 2.5 & 7,884,000.00 & 0.9 \\
4 & 1.25 & 3,942,000.00 & 0.9 \\
5 & 0.625 & 1,971,000.00 & 1 \\
6 & 0.3125 & 985,500.00 & 1 \\
7 & 0.15625 & 492,750.00 & 1 \\
8 & 0.078125 & 246,375.00 & 1 \\
9 & 0.0390625 & 123,187.50 & 1 \\
10 & 0.01953125 & 61,593.75 & 1 \\
\hline
\end{tabularx}
\caption{Emission Schedule \& Currency Supply}
\label{tab:emissions}
\end{table}
\FloatBarrier

This strategy ensures a gradual decrease in new currency introduction, maintaining mining incentives, and encouraging long-term investment. The model is designed for predictability and stability, supporting a balanced ecosystem. Future studies will examine its impact on market dynamics and network health.

\subsubsection{\$BASED Staking and Performance Incentives}
The incentive mechanism for users to earn \$BASED in the BasedAI ecosystem is by time released staking. Any user can stake \$BASED to any Brain or to any specific validator, including one they own. Users earn staking rewards from the emissions allocated to the validator they are staked to. To begin staking, users select a "Brain" (or associated validator) and delegate their staked amount to that Brain. To become a validator or a miner you must run a permanent memory operation to have that Brain “memorize” your validator or miner address.

\subsection{Initial Allocation Scheme}
The primary allocation of network-generated rewards, or 'emissions', in Brains adheres to a predefined ratio: 75\% to the nodes — the miners and validators — performing computational duties and the remaining 25\% to the Brain owner, or to the specific Brain validator entrusted with the staker's resources. This allocation scheme incentivizes the continuous oversight and optimized functioning of each Brain through the following optimizations:

\subsubsection{Owner Adjustments}
Brain owners maintain the discretion to enact redistribution of these emissions, this acts as an economic lever that will likely create a competitive environment for rewarding strong miners as the intelligence marketplace matures on BasedAI.

\subsubsection{Stake-Weighted Incentives}
Subsequently, the model incentives are magnified for users proportionately to their respective stakes within a Brain, creating a preferential structure that rewards more heavily-engaged participants — an attempt to equitably reward the relative commitment to the network but also the relative commitment to a given Brain.

\subsubsection{Validator Diversification Incentive}
\FloatBarrier
\begin{figure}[ht]
\centering
\begin{tikzpicture}
\begin{axis}[
    ybar,
    ymin=0,
    ymax=100,
    ylabel={Stake Amount},
    symbolic x coords={Brain 1, Brain 2, Brain 3},
    xtick=data,
    nodes near coords,
    nodes near coords align={vertical},
    legend style={at={(0.5,-0.2)},anchor=north,legend columns=-1},
]
\addplot+[ybar, fill=blue!40, opacity=0.6] coordinates {(Brain 1,80) (Brain 2,90) (Brain 3,85)};
\addlegendentry{Active Validators}

\addplot+[ybar, fill=red!40, opacity=0.6] coordinates {(Brain 1,60) (Brain 2,50) (Brain 3,40)};
\addlegendentry{Inactive Validators}

\draw [dashed, thick] (rel axis cs:0,0.7) -- (rel axis cs:1,0.7) node [pos=0.5, above] {70th Percentile};
\end{axis}
\end{tikzpicture}
\caption{Comparison of Validator Incentives}
\label{fig:validatorIncentives}
\end{figure}
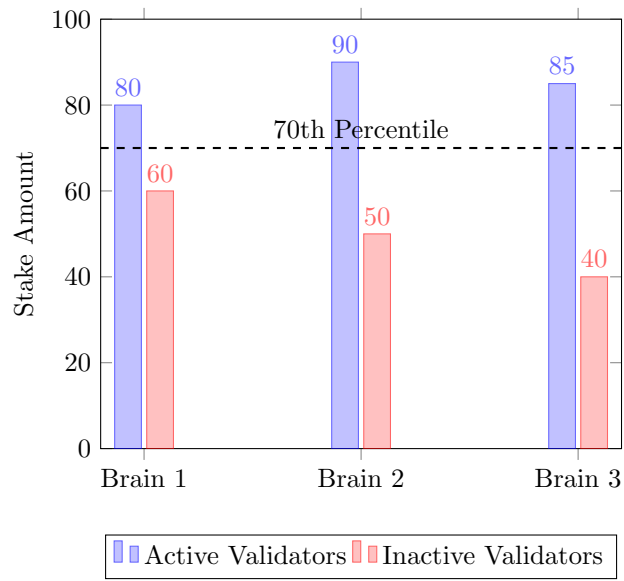
\FloatBarrier

BasedAI introduces a strategic delineation for active stakes for innovation amongst Brains, where only those validators with a stake amount in the top 70th percentile, in comparison to their counterparts within the Brain, are recognized as 'active'. This creates a dispersion dynamic that encourages validators to distribute their efforts (and subsequently stakes) across multiple Brains. Such a design is particularly favorable toward emerging Brains where initial stakes are characteristically lower.

\subsection{Example Fiscal Yield}

For illustrative purposes, let us consider a validator staking 100 \$BASED on Brain \#2. This validator cannot replicate the same stake on Brain \#11 — each Brain requires a distinct stake. Under this system, it is projected that validators might achieve an Annual Percentage Yield (APY) ranging between 15\% to 26\% in the form of \$BASED, a yield contingent upon the Brain's performance and the specific incentive configuration by the Brain owner.

\subsubsection{Median Emission Expectations}

From a broader perspective, the top 50\% of Brains can anticipate approximately 19\% APY on average. Extrapolated over an annual period, a 10,000 \$BASED stake in a performing Brain could manifest as 11,900 \$BASED, accruing an emission of 1,900 \$BASED.

\subsubsection{Mitigation of Centralization Risks}

To hinder the potential centralization akin to oligarchic dominance, a cap is instituted on Brain rewards, limiting allowable stakes to a maximum of 0.5\% of the aggregate network stake in \$BASED. This measure disperses the incentive fairly while allowing for a certain level of accessibility in Brain ownership and validating activities while curtailing the propensity to stake disproportionately in any single Brain.

\subsubsection{Emission Function, Intelligence Rewarding Intelligence}
To begin, it is necessary to refactor ``intelligence'' to be compatible with machine learning as a parameterized function capable of making predictions or decisions based on input data and its value to the network. It is important to note this is not a quantized FHE-compliant function. In BasedAI this function is managed by a Temporal Fusion Transformer (TFT) Enforcer (Lim, B. et al. 2021 \cite{lim2021}), which is trained over the temporal data of the chain $D=[X,Y]$ to minimize loss $L=E_D[Q(y,\text{TFT}(x))]$. TFTs excel at capturing complex temporal relationships within multivariate time-series data, such as transactions, block times, bytes of data, active address, emission rewards, etc...

The BasedAI network comprises $n$ peer functions $F=\text{TFT}_0,...,\text{TFT}_j,...,\text{TFT}_n$, each representing a validator entity within the BasedAI network that holds a stake $S=[s_i]$ recorded on a digital ledger. These entities contribute to a collective machine learning objective weighted by their respective stakes:
\[
\sum(L_i \cdot s_i)
\]
The goal is to distribute incentives $I$ proportionally to peers who effectively minimize the overall loss-objective and ultimately transition this distribution to be managed by the TFT Enforcer model. This distribution mechanism must be resistant to collusion and manipulation.

\subsection{Temporal Fusion Transformers (TFT) in BasedAI}

\subsubsection{TFT Enforcer In Consensus}
Peers use outputs from other peers' TFT Enforcers ($f(\text{TFT}(x))$) as inputs for their own models and learn inter-peer weights $W=[w_{i,j}]$ through transactions recorded on the digital ledger. Weights are set using Fisher's information pruning score (subject to variability injected by the TFT Enforcer) to reflect each peer's contribution towards reducing entropy when removed from the network.

To address potential collusion where peers might vote for themselves, we introduce an 'incentive' function $I(W,S,\text{TFT},t)$ that limits rewards to peers not reaching consensus in the network. The incentive mechanism requires both a stake vector $S$ and weight matrix $W$ for inter-peer rankings. A trust matrix $T$ is inferred from $W$, indicating non-zero edges between peers $i$ and $j$ if there is mutual ranking.
\[
W = [\ldots]
\]
\[
S = [\ldots]
\]
\[
T = [\ldots]
\]
Peers reach 'consensus' when they have non-zero edges from more than 50 percent of stake in the network $(T^T \cdot S) > 0.5$. To ensure differentiability, we employ a sigmoid function that scales rewards for connected peers while penalizing untrusted ones.

\subsubsection{Transition period for the TFT Enforcer}
During the first two years $(t \leq 2)$ calculated at a rate of $3,155,760$ blocks per year from the Genesis Block, the weighting factor $\beta(t) = 0$, meaning TFT predictions do not influence incentives. Over the next two years $(2 < t \leq 4)$, $\beta(t)$ linearly increases until it reaches 90\%, reflecting growing confidence in TFT accuracy. After four years $(t > 4)$, $\beta(t) = 1$, fully integrating TFT predictions into incentive calculations.
\[
I(W,S,\text{TFT},t) = R \cdot C + \beta(t) \cdot \text{Accuracy}(\text{TFT})
\]
This gradual approach allows stakeholders to adapt to the modeling capabilities provided by the TFT Enforcer while maintaining stability during implementation. In addition, the gradual impact of the TFT allows the results of different samples of metadata to be injected into the TFT if the impact is not improving accuracy or performance. The state covariate encoders can also be easily reversed so that the impact of the TFT can be undone if the impact is considered detrimental or less effective than the fixed approach.

If successful, the TFT Enforcer will enable BasedAI to dynamically reward Brains through its mechanisms. Assuming the management of rewards is successful, Based Labs plans to continue the network's evolution. Based Labs intends to deploy another version of the TFT Enforcer in Brain \#2, incorporating further optimizations. These optimizations include, but are not limited to, block time, block size, transaction routing, and consensus thresholds.

\subsubsection{Peers Trust Over Time Credits}
A peer is defined as a machine or machines connecting to the BasedAI through an IPv4 / IPv6 address and computekey.

The variability of peers joining and exiting the network is considered highly valuable training data. Within the architecture of the BasedAI network, there is a novel reward construct termed Credits. These serve as speculative weights of trust and endorsement that peers allocate to one another in recognition of their predictive acumen or substantive contributions to the collective intelligence. This is done simply by peers running validators who automatically reach miners or validators from outside their Brain.

The operational dynamics of Credits are delineated as follows:
\begin{itemize}
    \item Each peer possesses an intrinsic level of influence within the network, denoted by its stake $S$. Concurrently, peers evaluate each other's performance through a weight matrix $W$, which encapsulates the degree of mutual confidence.
    \item The weight matrix $W$ functions akin to an evaluative ledger, recording the perceived value each peer contributes to the network's predictive capabilities.
    \item Periodic recalibrations of rewards entail a computation that factors in these evaluative scores ($W$) alongside the existing stakes ($S$).
    \[
    \text{Credits} = [\ldots]
    \]
    \[
    \Delta \text{Credits} =W \cdot S
    \]
    \item In this schema, if Peer A assigns high evaluative scores to Peer B, then Peer A accrues additional Credits contingent upon Peer B's standing within the broader consensus. This mechanism incentivizes foresight in recognizing and supporting valuable contributors early on.
    \item As Credits accumulate, they confer augmented influence upon their holders, signifying a track record of judicious endorsements within the network.
\end{itemize}

\subsubsection{Securing BASED Block Rewards}
The overarching objective of integrating Temporal Fusion Transformer (TFT) Enforcer and building trust is to effectively allocate the BASED block reward to the most credible and contributive validators and miners. To address potential discrepancies or strategic manipulations within this system, BasedAI employs adaptive weighting:
\begin{itemize}
    \item Continuous analysis powered by the TFT Enforcer modal scrutinizes prediction veracity against actual outcomes, ensuring that theoretical models align with empirical realities.
    \item Should any divergence between Credit distribution of trust and genuine contributions be detected the weights ($W$) modulate automatically.
\end{itemize}

The phased integration of TFTs into the BasedAI network, rewarding BASED effectively, and implementing a sophisticated dynamic “Credit” distribution system of trust—BasedAI becomes an intelligent system with emergent capabilities and optimizations.

\subsection{\$BASED Utility}

In addition to rewarding miners and validators for their participation, \$BASED utility extends to the core functionalities of the network.

\subsubsection{Compute Units and Brain Operational Fees}

Compute Units quantify computational resource consumption for operating Large Language Models (LLMs) in the BasedAI ecosystem. They are purchased exclusively with \$BASED to link network incentives to computational services. The following information outlines the operational costs associated with the use of popular LLMs, denominated in \$BASED. It is to be used as an illustrative guide for how Brain owners and Miners may evaluate their pricing and demand for activities on the BasedAI network.
While Prompt Cost and Completion Cost indicate the amount of \$BASED needed to process a prompt and generate a response, respectively, Context provides the maximum context size that can be handled in a single operation, expressed in compute units. Figures represented assume 1 \$BASED = \$10 USD:

\FloatBarrier
\begin{table}[H]
\small
\centering
\begin{tabularx}{\textwidth}{
    |>{\hsize=1.5\hsize}X|
    >{\hsize=0.5\hsize\raggedleft\arraybackslash}X|
    >{\hsize=0.5\hsize\raggedleft\arraybackslash}X|
    >{\hsize=0.5\hsize\raggedleft\arraybackslash}X|
}
\hline
\thead{Model Name} & \thead{Prompt Cost \\ (\$BASED per 1k \\ compute units)} & \thead{Completion Cost \\ (\$BASED per 1k \\ compute units)} & \thead{Context \\ (compute units)} \\
\hline
OpenAI: GPT-4 & 0.03 & 0.06 & 8191 \\
OpenAI: GPT-3.5 Turbo & 0.01 & 0.02 & 4095 \\
Google: PaLM 2 Chat & 0.00025 & 0.0005 & 36864 \\
Anthropic: Claude v2 & 0.008 & 0.024 & 200000 \\
Mistral: Medium & 0.002778 & 0.008333 & 32000 \\
Nous: Hermes 2 Mistral 8x7B DPO & 0.00018 & 0.00054 & 8192 \\
Meta: Llama v2 70B Chat & 0.0007 & 0.0009 & 4096 \\
Perplexity: PPLX 70B Chat & 0.0007 & 0.0028 & 4096 \\
Goliath 120B & 0.0125 & 0.0125 & 6144 \\
Synthia 70B & 0.005 & 0.005 & 8192 \\
\hline
\end{tabularx}
\caption{Estimated Requirements for Popular Large Language Models (Source: OpenRouter, 2024)}
\label{tab:modelEstimates}
\end{table}
\FloatBarrier

All functions on the Based network carry a computational cost reflected in terms of \$BASED. Fees on the Based network can only be paid in the network’s native token, \$BASED.
Payments of \$BASED can be facilitated through spending accounts that authorize \$BASED allocations for activities like prompt submissions, data training, and AI/ML tasks on behalf of users.

The economic framework between \$BASED tokens and compute units facilitates a balanced interaction between the supply of and demand for computational services. This arrangement not only underscores the utility of \$BASED in the network, but also highlights BasedAI's ability to meet diverse computational demands efficiently.

\subsubsection{Network Participation and EVM Fees}

Miners or validators wishing to contribute their resources to a given Brain are required to pay a registration fee to the Brain's owner. While that fee is initially defaulted to 100 \$BASED, it can be adjusted by the Brain owner. This adjustment can be done at any time, affecting all subsequent contributors but grandfathering-in already registered contributors.

\begin{figure}[h]
  \centering
  \includegraphics[width=.75\linewidth]{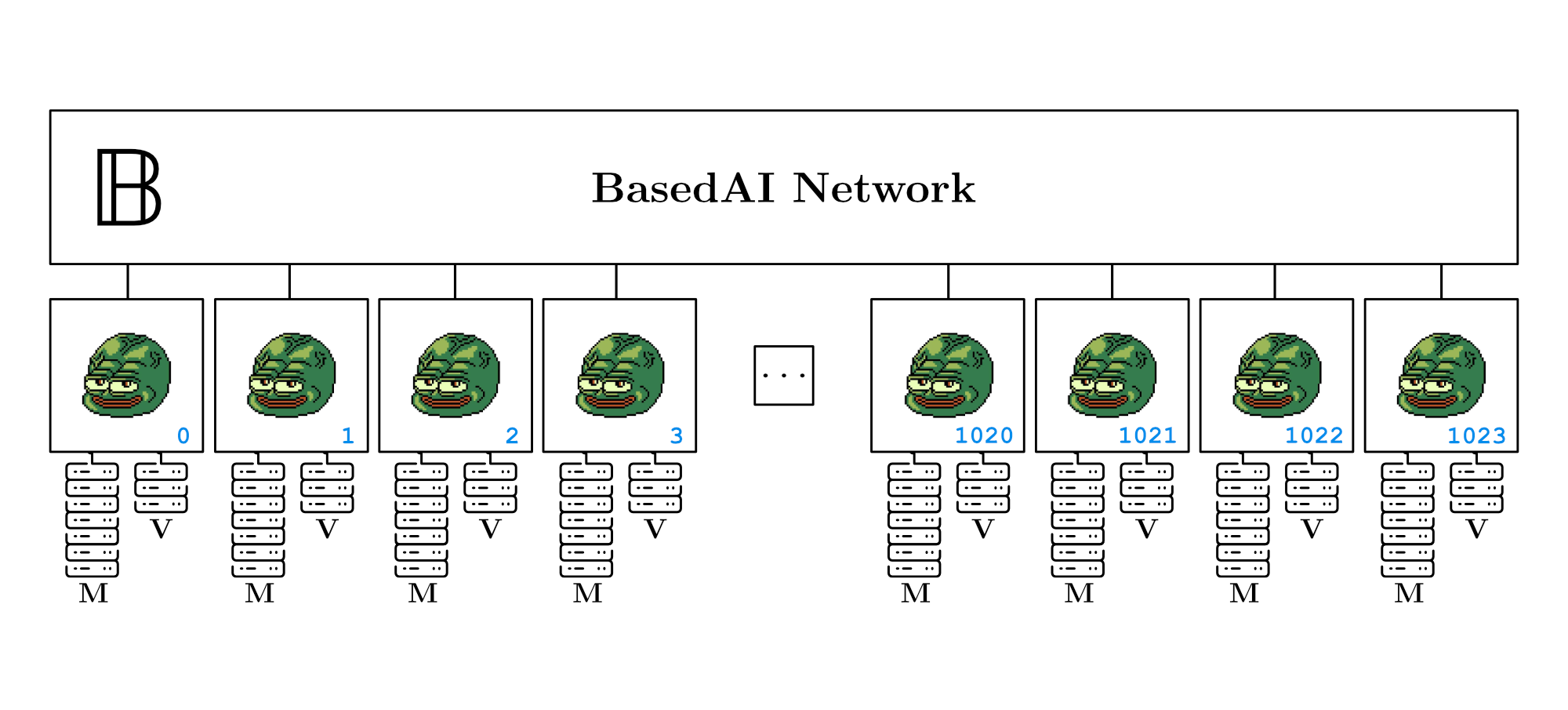}
  \caption{Brains, Miners, and Validators}
  \label{fig:basedDiagram}
\end{figure}

Each of the 1024 Brains in BasedAI’s network is capable of accommodating up to 256 validators (V) and 1792 miners (M). It is anticipated that a majority of Brains will not use all validator and miner slots in early stages of the network, however all bBains have the same max capacity.

BasedAI has an Ethereum Virtual Machine (EVM) compiler built-in. As a result, DeFi applications supported by an EVM are also supported in the BasedAI ecosystem. These services include Based Smart Contracts, Liquidity Markets, Based Swap, Based Pools, Kek Bot, and more.

\subsection{Governance and GigaBrains}

Pepecoin is an ERC-20 token used to fairly distribute Brains, which earn \$BASED emissions from every block created in the BasedAI network. A Brain is owned and managed through a tethered non-fungible token on Ethereum using the ERC-721 standard.

Brains that reach 0.5\% of the total stake of the network automatically become “GigaBrains” and can vote on network-critical decisions. 1 GigaBrain Vote = 1 GigaBrain Vote. Once Brains reach the GigaBrain threshold, they are allotted 1 vote. Increasing a Brain’s percentage of total stake will not allot that specific Brain more votes. Users may own more than one GigaBrain. Owners of Pepecoin are also permitted to participate in the governance of the network.

\FloatBarrier

\begin{tikzpicture}[
    node distance=2.5cm,
    auto,
    block/.style={
      rectangle,
      draw,
      fill=blue!20,
      text width=5em,
      align=center,
      rounded corners,
      minimum height=3em
    },
    line/.style={
      draw,
      -latex',
      thick,
      shorten >=2pt
    },
    decision/.style={
      diamond,
      aspect=2,
      draw,
      fill=red!20
    }
  ]

  \node[block] (pepecoin) {Pepecoin (ERC-20)};
  \node[block, below of=pepecoin] (brains) {Brains (ERC-721)};
  \node[block, below of=brains] (gigabrains) {GigaBrains};
  \node[decision, below of=gigabrains] (vote) {1 GigaBrain Vote};
  \node[block, right=3cm of vote] (gigabrainvote) {GigaBrain Voting Process};
  \node[block, left=3cm of vote] (basedlabs) {Based Labs Leadership};
  \node[block, below of=vote, text width=6em] (governance) {Network Governance};

  \draw[line] (pepecoin) -- (brains) node[midway] {acquires};
  \draw[line] (brains) -- (gigabrains) node[midway] {w/0.5\% of stake};
  \draw[line] (gigabrains) -- (vote);
  \draw[line] (vote) -- (gigabrainvote) node[midway] {exerts};
  \draw[line] (vote) -- (basedlabs) node[midway] {proposes};
  \draw[line] (pepecoin.west) -| (basedlabs.north) node[near start, above] {influences};
  \draw[line] (basedlabs.south) |- (governance.west) node[near end, below] {oversees};
  \draw[line] (gigabrainvote) |- (governance.east) node[near end, below] {influences};

\end{tikzpicture}

\FloatBarrier

 In the first year after the Genesis Block, the leadership at Based Labs will have a considerable role in network governance, with a system where proposals pass by default unless voted down by a majority. Brains will have a light security audit where owners must share git repository links for submitted Brain tasks, providing miners and validators with added confidence.

\section{Cerberus Squeezing and Model Optimization}
\subsection{Cerberus Squeezing Explained}

\begin{figure}[h]
  \centering
  \includegraphics[width=.75\linewidth]{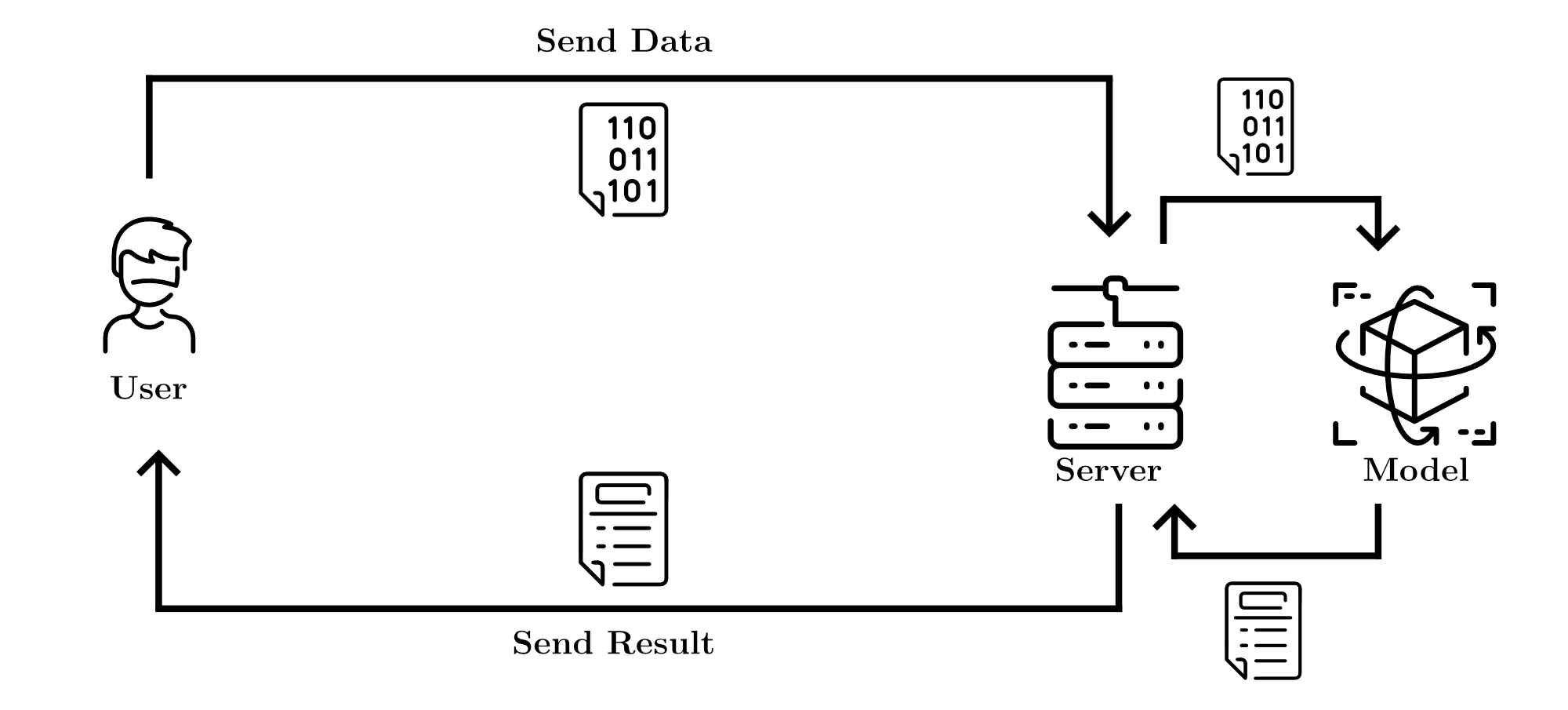}
  \caption{An Unencrypted Model}
  \label{fig:unencryptedModel}
\end{figure}

Large Language Models (LLMs), particularly those from the GPT series, have contributed substantially to progress in artificial intelligence. Yet, the commonly employed architecture for public-facing LLMs, as depicted in Figure 2, inherently presents several privacy risks. These include, but are not limited to, potential vulnerabilities from unencrypted data transfer, compromised servers, and the unauthorized surveillance of model interactions and outputs.

Fully Homomorphic Encryption (FHE) represents a significant step forward for addressing privacy concerns within computational frameworks. FHE’s capability to process encrypted inputs and return encrypted outputs ensures that data remains secure throughout the entirety of the computational process, accessible only to users with the appropriate decryption keys. This secure approach to computation allows for the extensive use of encrypted data without the risk of exposure or unauthorized access.

\begin{figure}[H]
  \centering
  \includegraphics[width=.75\linewidth]{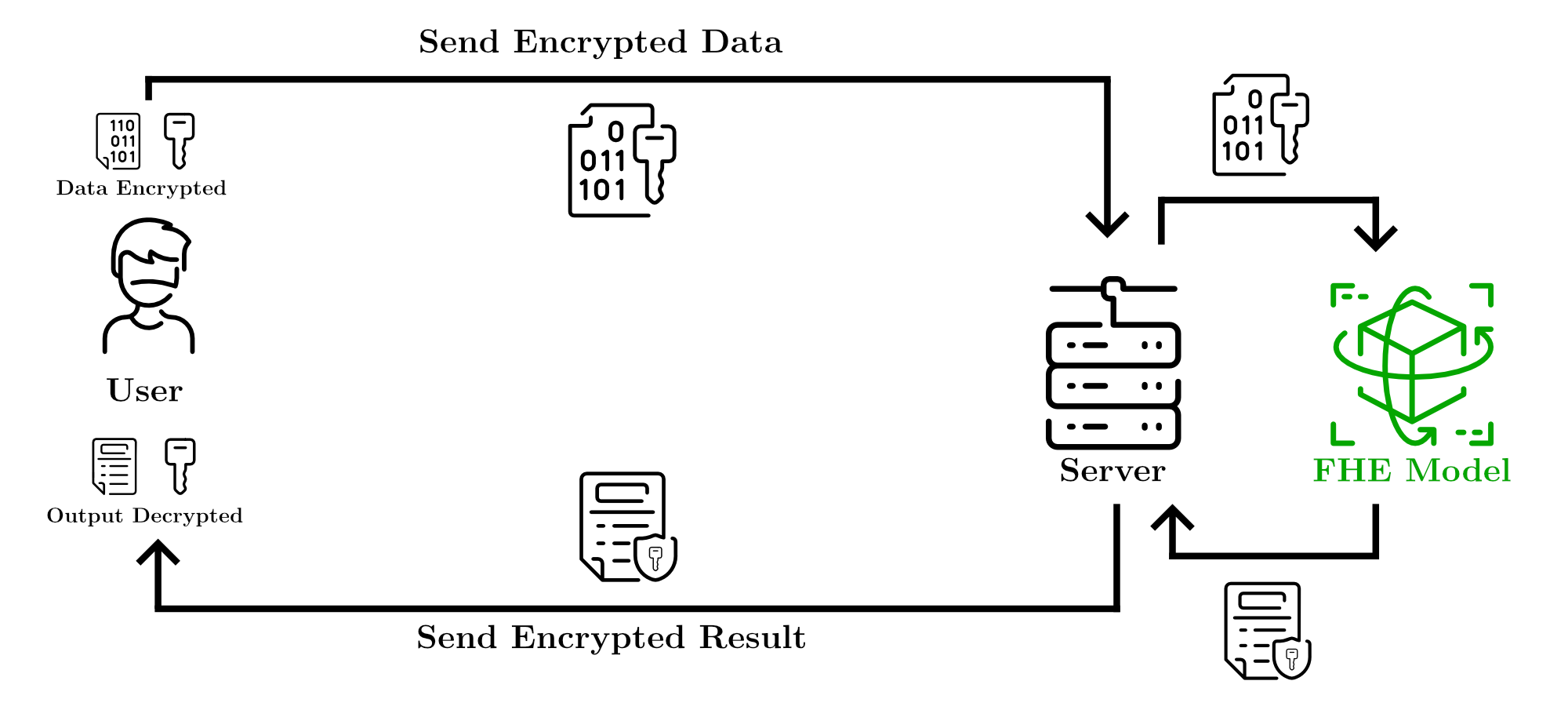}
  \caption{An FHE-Compliant Model}
  \label{fig:fheCompliantModel}
\end{figure}

As pictured above, the integration of FHE into LLMs is a promising concept that enhances the privacy to AI/ML systems and leverages zero-knowledge proofs to do so, drawing on foundational concepts such as those outlined by Ben-Sasson et al. (2014 \cite{sasson2014}). However, the integration of Fully Homomorphic Encryption into LLMs is not without tradeoffs. Unlike traditional encryption methods, which typically encrypt and decrypt data in bulk, FHE works on encrypted data at the bit level, allowing users to compute directly on ciphertexts. This means that every operation, no matter how minor, involves complex mathematical transformations that deal with polynomials or integer arithmetic under encryption.  This process is referred to as quantization (Han, S. et al. 2015 \cite{han2015}).

Furthermore, FHE requires frequent adjustments to the 'noise' built into the encryption to maintain security across operations. Every calculation generates additional noise, which must be carefully controlled to prevent decryption errors. Considering that even simple operations become polynomial calculations and one function might entail thousands to millions of such operations, the computational intensity quickly multiplies. For example, consider a simple expression like below:

\[
(a + b) + (c \times d) + (e \times f)
\]

When using a standard encryption process, the number of computational actions necessary is simply one (1): encrypt the whole function. Conversely, the FHE process necessitates that each bit is encrypted, in this simplified example that means each variable needs to be encrypted (6) and each function needs to be encrypted (5).

\begin{figure}[ht]
\centering
\begin{tikzpicture}
\begin{axis}[
    ybar,
    bar width=25pt,
    enlarge x limits=0.5,
    ymin=0, ymax=12,
    ylabel={\# of Computational Steps},
    xtick={1,2},
    xticklabels={Standard Encryption, FHE},
]

\addplot[fill=blue!40, fill opacity=0.6] coordinates {(1,1)};

\addplot[fill=red!40, fill opacity=0.6] coordinates {(2,11)};

\end{axis}
\end{tikzpicture}
\caption{Comparison of Standard Encryption vs. Fully Homomorphic Encryption (FHE)}
\label{fig:simpleFunctionsFHE}
\end{figure}
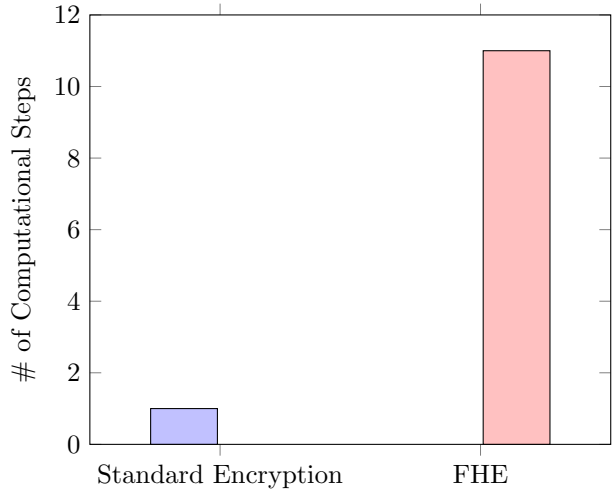

With the need to manipulate and manage noise levels in addition to the already complex calculations, FHE demands an immense amount of computational power, much more than unencrypted or traditionally encrypted computations. As a result, FHE for LLMs leads to significantly longer processing times, increased costs due to higher energy consumption, and a demand for more powerful or specialized computing hardware to maintain reasonable performance levels.

\subsection{Optimizing FHE Performance with Cerberus Squeezing}

To address the computational burden posed by FHE, BasedAI uses Cerberus Squeezing, a technique that applies principles of deep compression, significantly enhancing the efficiency of LLM operations (Vaswani, A. 2017 \cite{vaswani2017}). Cerberus Squeezing represents a targeted refinement for FHE within the realm of LLMs, aiming to enhance data privacy as well as computational efficiency in distributed networks. By selecting additional layers in the quantization process within LLM architectures and establishing dedicated Fully Homomorphic Encryption (FHE) circuits for these layers, the approach ensures that data remains encrypted throughout the processing phase and doesn't break the computational bank. This optimization of computational functions within the circuits significantly reduces the processing load for BasedAI miners while facilitating the secure and efficient operation on encrypted data. Consider the simple equation from earlier:

\[
(a + b) + (c \times d) + (e \times f)
\]

In an FHE environment without optimization, each variable requires distinct encryption, resulting in (6) encryption actions, and each arithmetic operation is also encrypted, totaling (5) additional steps, leading to (11) actions.

In contrast, when we apply Cerberus Squeezing, the number of computations is dramatically reduced. Rather than encrypting step by step, Cerberus Squeezing allows multiple operations to be merged into a single encrypted computation. This optimization can be conceptually likened to compressing multiple arithmetic tasks into a more efficient format that is naturally conducive to FHE. By restructuring data and computations, Cerberus Squeezing minimizes the encryption actions required for the same expression without sacrificing privacy. In this example, the functions that are used on each of the variables can be merged into a single encrypted computation with Cerberus Squeezing. This results in a more manageable and significantly reduced number of steps, reducing the (11) actions seen previously to just (5) actions in this simplified example.

\FloatBarrier
\begin{figure}[ht]
\centering
\begin{tikzpicture}
\begin{axis}[
    ybar,
    bar width=25pt,
    enlarge x limits=0.5,
    ymin=0, ymax=12,
    ylabel={\# of Computational Steps},
    xtick={1,2},
    xticklabels={FHE, FHE w/Cerberus Squeezing},
]

\addplot[fill=blue!40, fill opacity=0.6] coordinates {(1,11)};

\addplot[fill=red!40, fill opacity=0.6] coordinates {(2,5)};

\end{axis}
\end{tikzpicture}
\caption{Comparison FHE without Optimization vs. FHE with Cerberus Squeezing}
\label{fig:cerberusSqueezingFHE}
\end{figure}
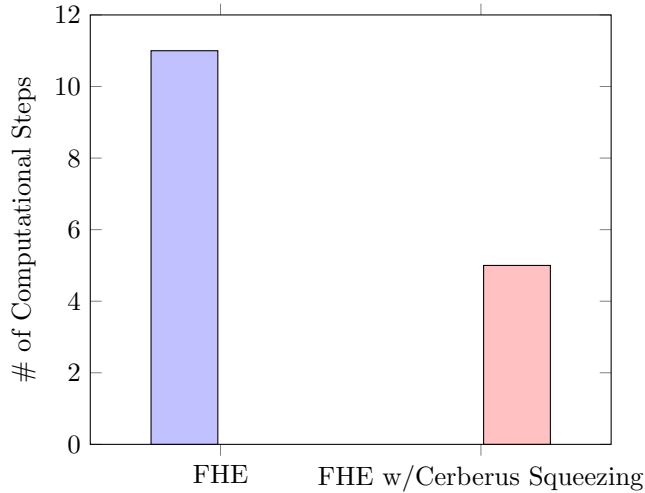
\FloatBarrier

While in this example, optimizing the quantization of FHE for isolating mathematical functions to increase efficiency is instructive in understanding the core concepts, Cerberus Squeezing when applied to ZK-LLMs is much more complex. By choosing to optimize the multihead attention mechanism (MHA) found in advanced transformer models like GPT-2+, BasedAI ensures the GPT models operate within a specified computational budget. The Cerberus Squeezing method modifies the GPT model’s forward pass to handle FHE operations without revealing content before it is received by network participants.
\begin{figure}[h]
  \centering
  \includegraphics[width=.35\linewidth]{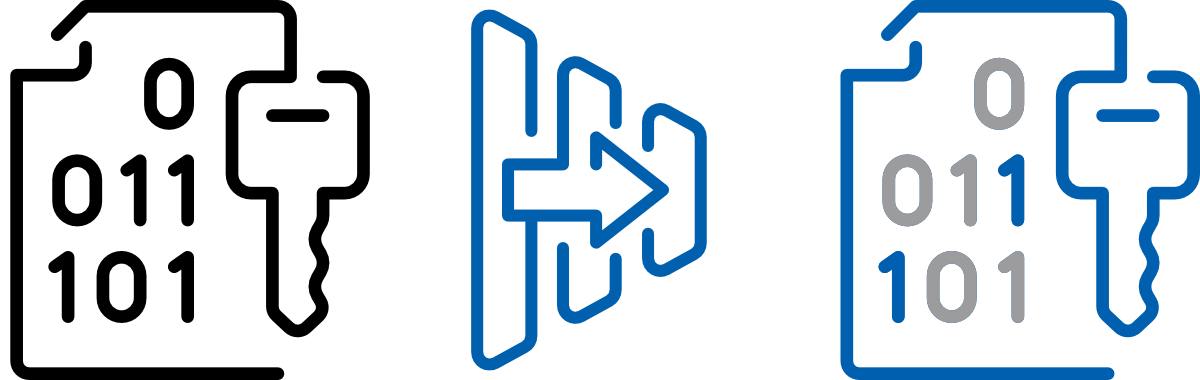}
  \caption{Cerberus Squeezing}
  \label{fig:cerberusDiagram}
\end{figure}

Additionally, because the Cerberus Squeezing optimization is specific to the MHA mechanism found in most transformer models (Jacob, B. et al. 2018 \cite{jacob2018}), any model that relies on transformers, whether existing models (Llama) or future models (GPT-5) are compatible with BasedAI and can be transformed into a ZK-LLM on the network.

\subsection{Optimizing Cerberus Squeezing Over The Network}

To improve the performance of FHE the core of the Cerberus Squeezing is optimized when the Brain routes the tasks for processing to the miners. By default, the embeddings are sorted on the miner and on the nodes of the Brain distributed work

\vspace{\baselineskip}
\begin{lstlisting}[
  basicstyle=\scriptsize\ttfamily,
  backgroundcolor=\color{white},
  showspaces=false,
  showstringspaces=false,
  showtabs=false,
  frame=single,
  tabsize=2,
  captionpos=b,
  breaklines=true,
  breakatwhitespace=false,
  escapeinside={\%*}{*)},
  language=Python, title={\bfseries Embedding Radius Calculation}]
import torch
def calculate_embedding_radius(current_embedding, previous_embeddings):
    # Stack all previous embeddings into a single tensor for batch processing
    prev_embeddings_tensor = torch.stack(previous_embeddings)
    # Compute distances (e.g., Euclidean distance)
    distances = torch.norm(prev_embeddings_tensor - current_embedding.unsqueeze(0), dim=1)
    return distances
\end{lstlisting}

This resulting radius is grouped into a covariance matrix and distributed using the matrix.

\vspace{\baselineskip}
\begin{lstlisting}[
  basicstyle=\scriptsize\ttfamily,
  backgroundcolor=\color{white},
  showspaces=false,
  showstringspaces=false,
  showtabs=false,
  frame=single,
  tabsize=2,
  captionpos=b,
  breaklines=true,
  breakatwhitespace=false,
  escapeinside={\%*}{*)},
  language=Python,  title={\bfseries Work Distribution to Miners}]
def distribute_work_to_miners(data_batch, local_embedding, peer_nodes):
    work_assignments = {}
    for data_item in data_batch:
        # Assume we have a function that generates an embedding for this piece of data
        item_embedding = generate_embedding_for_data(data_item)
        # Calculate radii to all peers' previous embeddings
        radii_to_peers = {peer_id: calculate_embedding_radius(item_embedding, peer_embeddings)
                          for peer_id, peer_embeddings in peer_nodes.items()}
        # Find the closest peer based on minimum radius
        closest_peer = min(radii_to_peers, key=radii_to_peers.get)

        # Assign this piece of data to the closest peer
        if closest_peer not in work_assignments:
            work_assignments[closest_peer] = []
        work_assignments[closest_peer].append(data_item)
    return work_assignments
\end{lstlisting}

The result allows the miners to leverage automated oversubscription optimization on NVIDIA GPUs as well as the vector databases using access to organize memory allocation.

\section{Implementing Dynamic Quantization}

Cerberus Squeezing integrates dynamic quantization with adaptive scaling to optimize data preprocessing, striking a balance between computational efficiency and data privacy and drawing upon Krishnamoorthi's insights on quantizing deep convolutional networks (2018 \cite{krishnamoorthi2018}). Below, we detail the mathematical algorithm that underpins this process, followed by its pseudocode implementation to illustrate its practical application alongside a GPT-2 model.

\subsection{Mathematical Algorithm for Dynamic Quantization}

The dynamic quantization process is designed to adjust the precision of input data dynamically, based on its variability, to optimize for computational resources while maintaining the integrity of the data's informational content. This process involves three key steps:

\subsubsection{Standard Deviation Calculation}
For an input tensor \( X \) with dimensions \( N \times M \), calculate the standard deviation \( \sigma_i \) for each sample \( i \), where \( N \) represents the number of samples and \( M \) the feature dimension.

\[
\sigma_i = \sqrt{\frac{1}{M}\sum_{j=1}^{M}(x_{ij} - \mu_i)^2}
\]

\subsubsection{Adaptive Scaling}
Apply an adaptive scaling factor \( S_i \) to each sample, based on its standard deviation \( \sigma_i \), using a predefined threshold \( T \) and scaling factor \( \alpha \).

\[
S_i =
\begin{cases}
\alpha & \text{if } \sigma_i > T \\
\frac{1}{\alpha} & \text{otherwise}
\end{cases}
\]

\subsubsection{Quantization}
Quantize the scaled inputs to a predetermined number of levels \( L \), by mapping the feature values of each sample to the nearest quantization level.

\[
q_{ij} = \left\lfloor \frac{x_{ij} - \text{min}_i}{Q_i} \right\rfloor \cdot Q_i + \text{min}_i
\]

This algorithm significantly simplifies the computational demands associated with processing encrypted data, enabling more efficient and scalable AI applications in secure environments.

\subsection{Pseudocode Implementation (Illustrative)}

Below is the pseudocode for the Cerberus Squeezing technique, demonstrating its application with a GPT-2 model for dynamic quantization and adaptive scaling. The technique is particularly designed to support ZK-LLMs, which leverage the distributed network of BasedAI to host the models and Fully Homomorphic Encryption (FHE) to encrypt the input and output throughout the lifecycle of data processing.
\vspace{\baselineskip}
\begin{lstlisting}[
  basicstyle=\scriptsize\ttfamily,
  backgroundcolor=\color{white},
  showspaces=false,
  showstringspaces=false,
  showtabs=false,
  frame=single,
  tabsize=2,
  captionpos=b,
  breaklines=true,
  breakatwhitespace=false,
  escapeinside={\%*}{*)},
  language=Python,  title={\bfseries Cerberus Squeezing Pseudocode}]
# SqueezingModule: Preprocess data via quantization & scaling
Module SqueezingModule:
  Initialize(quant_levels: 256, adap_thresh: 0.1, scale_factor: 0.5)

  # Forward: Apply scaling & quantize
  Forward(inputs):
    var = StdDev(inputs) # Compute input variability
    scaled = var > adap_thresh ? ScaleDown(inputs, scale_factor) : ScaleUp(inputs, scale_factor)
    quantized = Quantize(scaled, quant_levels) # Quantize scaled inputs
    Return quantized

# zkLLM: Integrates SqueezingModule with GPT
Model zkLLM:
  Initialize(gpt_model)

  # Forward: Preprocess & feed to GPT
  Forward(inputs):
    preprocessed = SqueezingModule.Forward(inputs) # Scale & quantize
    outputs = gpt_model.Process(preprocessed) # Process with GPT
    Return outputs

# Demo: Use zkLLM with a sample input
Main:
  gpt_model = LoadGPTModel("pretrained-GPT")
  tokenizer = LoadTokenizer("pretrained-GPT")

  zkLLMInstance = zkLLM(gpt_model) # Init ZkLLM

  input_text = "Hello, how are you?"
  tokenized = tokenizer.Tokenize(input_text) # Tokenize input

  output = zkLLMInstance.Forward(tokenized) # Process input
  Print(output) # Display output
\end{lstlisting}

This pseudocode underscores the Cerberus Squeezing technique's pivotal role in pre-processing inputs for language models run on BasedAI’s network. The SqueezingModule meticulously adjusts input data, optimizing it for both processing efficiency and data privacy, especially critical in handling encrypted information. By dynamically scaling and quantizing inputs, it simplifies the computational load, paving the way for more efficient encrypted data processing.

Furthermore, the zkLLM class embodies the integration of this optimized preprocessing with the GPT-2 model's computational linguistics capabilities. By modifying the model's default data handling pathway to incorporate the SqueezingModule, it guarantees that all data is primed for processing, spotlighting the technique's capacity to enhance AI application performance in secure, data-sensitive contexts. This approach not only boosts efficiency of processing encrypted data but also marks a noteworthy advancement in deploying large language models within privacy-focused applications.

\section{The Future of Privacy with ZK-LLMs}

By leveraging the advancements of cutting-edge encryption within a distributed network, BasedAI positions ZK-LLMs at the forefront of AI development, emphasizing both user privacy and computational efficiency. This section explores the practical implementations of ZK-LLMs, illustrating their versatility and transformative potential for industries prioritizing secure data handling.

\subsection{Example ZK-LLM Workflow: Healthcare}
Consider a hypothetical scenario involving a specialized Brain, dubbed the "Medical Records" Brain. This Brain is designed to process sensitive medical data on the BasedAI network, offering insights and recommendations without ever compromising patient privacy. Here’s how it would work:

\begin{figure}[h]
  \centering
  \includegraphics[width=.75\linewidth]{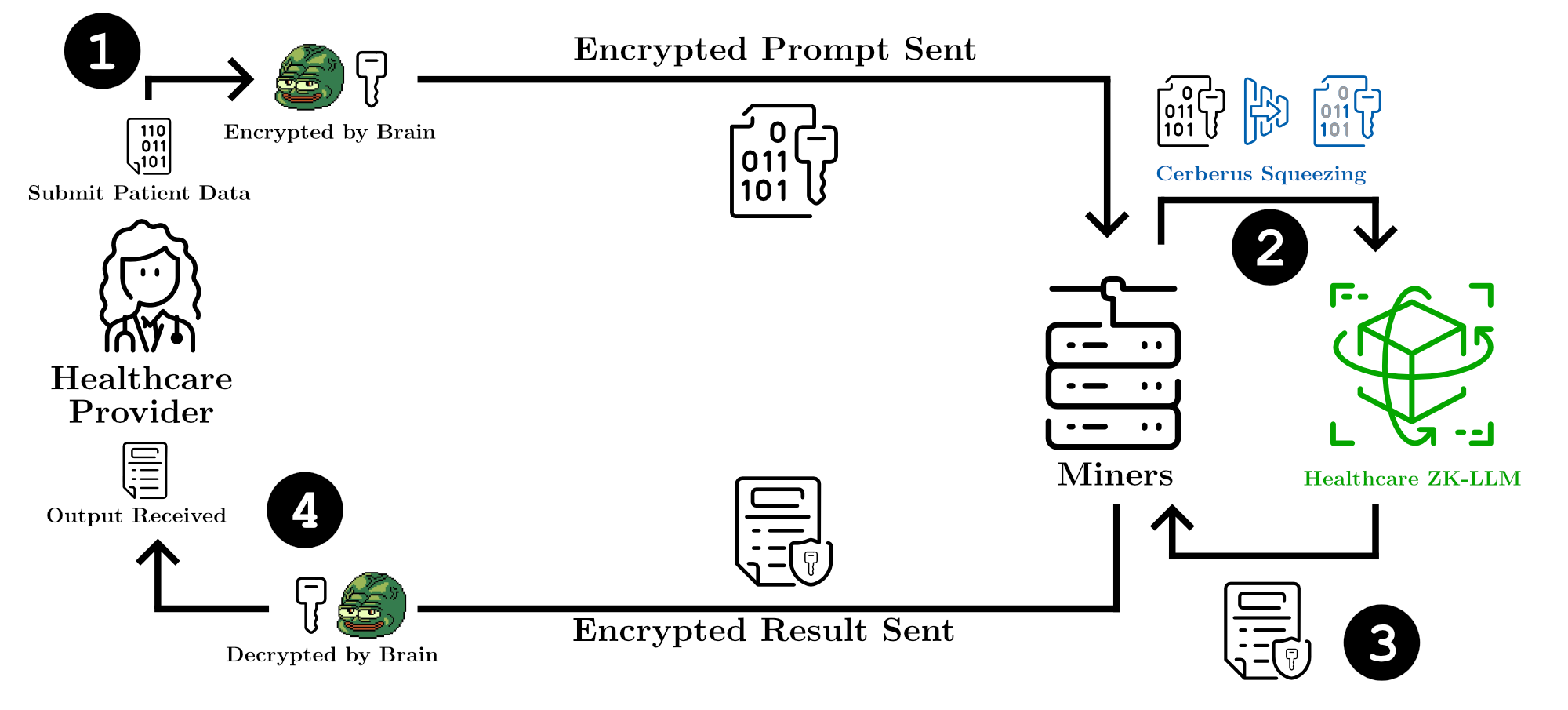}
  \caption{Healthcare workflow using a ``Medical Records" Brain}
  \label{fig:healthcareWorkflow}
\end{figure}

\begin{enumerate}
    \item Encrypted Query Submission to BasedAI: A healthcare provider sends encrypted patient data as a query to BasedAI's ``Medical Records" Brain to seek treatment insights. This query specifies the type of analysis or insight the healthcare provider is seeking, such as potential treatment options for a specific condition. The Brain runs a ZK-LLM which ensures data privacy upfront by encrypting the query itself, making it unreadable to unauthorized parties.

    \item Processing by Miners: Miners in the BasedAI network, who have allocated their computational resources to the ``Medical Records" Brain, receive the encrypted query. Utilizing the ``Cerberus Squeezing" process, these miners process the query on their encrypted ZK-LLMs. Importantly, the miners can perform complex computations required for analyzing the medical data without ever decrypting it, thus preserving privacy throughout the process.

    \item Generating Encrypted Insights: Once the analysis is complete, the ZK-LLM generates insights, such as recommended treatment plans, based on the encrypted medical records. It is important to note that these insights remain completely encrypted so that only the healthcare provider can decrypt and understand them.

    \item Receiving Encrypted Results: The healthcare provider receives the encrypted output from the BasedAI network. Only the submitting user has the ability to decrypt their results, gaining access to the treatment recommendations without the data ever having been exposed or compromised during the process.
\end{enumerate}

\subsection{Expanding ZK-LLM Applications}

BasedAI’s ZK-LLMs demonstrate significant promise in a wide variety of sectors where data privacy is pivotal. In each example application below, the capacity to handle sensitive data securely remains central:

\begin{enumerate}
    \item \textbf{Secure Financial Trading and Analysis:} A ``Financial Trading'' ZK-LLM could provide real-time, encrypted analysis of market data to offer trading insights and strategies without compromising sensitive financial information, catering to high-frequency trading firms and individual investors.

    \textit{Example:} Consider a financial information company such as Bloomberg, serving a global clientele with access to a wealth of market data. Such an organization may face challenges in processing sensitive data while adhering to various international data privacy laws. Deploying a Brain specifically for secure, encrypted processing of market data could enable the firm to offer its clients bespoke market insights that are fully compliant with global privacy standards, thus ensuring that strategic information remains protected throughout the analytical process.

    \item \textbf{Cybersecurity Threat Intelligence:} A ``Cybersecurity Intelligence'' ZK-LLM could process encrypted network traffic data to identify and predict cybersecurity threats in real-time, offering a high-value service to corporations and governments.

    \textit{Example:} An organization like CrowdStrike could benefit from using a Brain dedicated to cybersecurity threat detection into its infrastructure for the purpose of analyzing encrypted network traffic. This application would enhance their existing endpoint security solutions by adding another layer of privacy. By utilizing such a Brain, businesses would be able to fortify their cyber defenses and safeguard their digital assets against threats, while ensuring their sensitive internal network data remains confidential throughout the analysis process.

    \item \textbf{Anonymous Search:} A ``Private Search'' ZK-LLM could process search queries in an encrypted form, offering users the ability to search the internet without compromising their privacy. This application would be invaluable for privacy-focused search engines and web browsers seeking to enhance user confidentiality.

    \textit{Example:} Search engines focused on privacy such as Brave or DuckDuckGo, could integrate a Brain into their system to encrypt user search requests. Such a development would further reinforce the search engine's dedication to user privacy by mitigating the risks of tracking or profiling based on search behavior. By leveraging ZK-LLMs, users' queries could be processed in a secure environment that respects and protects their desire for anonymity.

    \item \textbf{Bittensor Subnet:} A ``Secure AI Network'' ZK-LLM could enable the encrypted sharing and processing of data across distributed AI networks, like Bittensor subnets. This facilitates secure, collaborative AI development and deployment, protecting the data's integrity and confidentiality.

    \textit{Example:} TensorFlow could expand its open-source machine learning framework capabilities by incorporating a Brain into its ecosystem. This addition would provide developers with the tools to collaborate on training and deploying machine learning models while preserving data privacy with ZK-LLMs. By using such an encrypted, distributed approach, TensorFlow could set a new standard for privacy-preserving AI development and operations.
\end{enumerate}

These applications significantly broaden the scope of ZK-LLMs, venturing into sectors where privacy is paramount while also exploring markets with substantial revenue opportunities. By leveraging the distinctive capabilities of BasedAI to deliver innovative and secure solutions, and ensuring that sensitive data remains encrypted from query submission through to the delivery of insights, BasedAI is forging a path towards a future where advanced AI applications can be implemented in sensitive domains without sacrificing privacy or security.

\section{Appendix}

BasedAI introduces Cerberus Squeezing as a method to enhance the practicality and performance of Fully Homomorphic Encryption (FHE) applications. This approach optimizes the efficiency and speed of neural network operations on encrypted data by focusing on computational resource allocation within multi-head attention mechanisms. Furthermore, BasedAI's research into ``Encrypted Data Synthesis" and ``Quantization-Aware Training" aims to address challenges such as limited data accessibility and the need for models to adapt to quantization effects, thereby expanding secure training resources and improving protocol performance.

\subsection{Cerberus Squeezing}
``Cerberus Squeezing" is a technique aimed at optimizing multi-head attention mechanisms in neural network models, particularly for processing encrypted data under FHE. This involves selectively focusing computational resources on the most impactful attention heads to enhance efficiency and performance.

\subsubsection{Multi-Head Attention Mechanism}
The multi-head attention mechanism in a neural network model can be represented as a function \(A\) that operates on an input sequence \(X\) to produce an output sequence \(Y\), where each head \(h\) focuses on different parts of \(X\):
\[
Y = A(X) = \text{Concat}(head\_1, head\_2, \ldots, head\_k)W_O
\]
\[
head\_i = \text{Attention}(XW^Q_i, XW^K_i, XW^V_i)
\]
where \(W^Q_i\), \(W^K_i\), and \(W^V_i\) are weight matrices for the i-th head’s query, key, and value, respectively, and \(W_O\) is the output weight matrix.

\subsubsection{Selective Optimization}
Cerberus Squeezing identifies and prioritizes heads that contribute most significantly to the model’s performance. If \(S(h)\) represents the significance score of head \(h\), computational resources are focused on heads with \(S(h) > \theta\), where \(\theta\) is a threshold determining head significance.

\subsubsection{Resource Allocation}
This optimization can be formalized as a constrained optimization problem, aiming to maximize model performance under computational resource constraints:
\[
\max_{W^Q_i, W^K_i, W^V_i} \text{Performance(Model)}
\]
subject to
\[
C(W^Q_i, W^K_i, W^V_i) < C_{\text{max}}
\]
where \(C(\cdot)\) measures the computational cost, and \(C_{\text{max}}\) represents the maximum allowable computational budget.

\subsection{Further Research}
\subsubsection{Encrypted Data Synthesis}
``Encrypted Data Synthesis'' refers to the process of generating synthetic data points from existing encrypted datasets without decrypting them. Applied to BasedAI, this technique would leverage the inherent patterns and structures within a given set of data, preserved under encryption, to fabricate new, realistic data points that can enhance BasedAI model training.

\textbf{Encrypted Data Representation} Consider an encrypted data point represented as \( E(x_i) \), where \( x_i \) is the original data point, and \( E(\cdot) \) denotes the encryption function. The dataset of encrypted points is denoted as \( D_{\text{enc}} = \{E(x_1), E(x_2), \ldots, E(x_n)\} \).

\textbf{Pattern Extraction} Assuming a function \( F_{\text{pattern}} \) that operates on encrypted data to identify patterns without decryption:
\[ F_{\text{pattern}}(D_{\text{enc}}) \rightarrow P_{\text{enc}}, \]
where \( P_{\text{enc}} \) represents the pattern information extracted from \( D_{\text{enc}} \), still in encrypted form.

\textbf{Data Fabrication} The fabrication of new data points, \( E(x_{\text{new}}) \), uses the pattern information \( P_{\text{enc}} \) to guide the generation process through a generative model \( G \):
\[ E(x_{\text{new}}) = G(P_{\text{enc}}), \]
where \( G \) is designed to work within the FHE scheme, ensuring that the output is a realistically fabricated data point, still encrypted.

\subsubsection{Quantization-Aware Training}
Quantization-Aware Training (QAT) optimizes neural network models for efficient deployment in environments with limited computational resources. This technique adjusts models to operate effectively with lower precision arithmetic, crucial for maintaining performance under FHE constraints.
\[
Q(x) = \text{round}\left(\frac{x - \mu}{\sigma}\right) \cdot q_{\text{scale}} + q_{\text{zero}}
\]
where \(x\) is the input, \(\mu\) and \(\sigma\) are the mean and standard deviation of the input distribution, \(q_{\text{scale}}\) is the scaling factor for quantization, and \(q_{\text{zero}}\) is the zero-point offset.

During QAT, the model is trained or fine-tuned with simulated quantization effects, integrating the quantization function into the forward pass. This allows the model to adapt to the quantized representation before actual deployment:
\[
F_{\text{QAT}}(x) = Q^{-1}(Q(x))
\]
where \(F_{\text{QAT}}\) represents the model's forward pass incorporating quantization, and \(Q^{-1}\) denotes the dequantization operation to simulate the effect of quantization on the model's performance.

The objective of QAT is to minimize the discrepancy between the performance of the quantized model and its full-precision counterpart, ensuring efficiency and accuracy in resource-constrained environments.

\end{document}